\documentclass[notitlepage,reqno,12pt]{article}
\usepackage[T1]{fontenc}
\usepackage[utf8]{inputenc}
\usepackage{lmodern}

\usepackage[affil-it]{authblk}
\usepackage{amsmath,amssymb,amsfonts,amsthm,amscd,mathrsfs}
\usepackage{mathtools}
\usepackage{geometry}
\usepackage{hyperref}

\usepackage{tikz}
\usetikzlibrary{matrix,arrows,decorations.pathmorphing,positioning, shapes, fit}
\usepackage{times}
\usepackage{mathbbol}
\usepackage{colonequals}
\usepackage{mathbbol}
\usepackage{colonequals}

\usepackage{graphicx}
\usepackage{color}

\usepackage{xcolor}
\usepackage{todonotes}

\usepackage[all]{xy}

\hypersetup{
  colorlinks = true,
  urlcolor = blue,
  linkcolor = blue,
  citecolor = red,
  pdfauthor = {Moutuou, E.},
  pdfkeywords = {},
  pdftitle = {Moutuou, E. - Graphs are focal hypergraphs},
  pdfsubject = {Complex networks},
  pdfpagemode = UseNone
}

\theoremstyle{definition}
\newtheorem{definition}{Definition}[section]

\theoremstyle{remark}
\newtheorem{remark}[definition]{Remark}

\theoremstyle{plain}
\newtheorem{proposition}[definition]{Proposition}
\newtheorem{lemma}[definition]{Lemma}

\newcommand{\N}[1]{N[#1]}

\newcommand{\R}{\mathbb{R}}
\newcommand{\calI}{\mathcal{I}}
\newcommand{\calP}{\mathcal{P}}
\newcommand{\calN}{\mathcal{N}}
\newcommand{\Nf}{\calN_{\mathrm f}}
\newcommand{\calH}{\mathcal{H}}
\newcommand{\calF}{\mathcal{F}}

\newcommand{\TwoSec}{\operatorname{P}}

\begin{document}

\title{Graphs as focal interaction systems in higher-order dynamics}

\author{Elka\"ioum M. Moutuou %
 \thanks{E-mail: elkaioum.moutuou@concordia.ca}}
 \affil{Department of Electrical and Computer Engineering \\
Concordia University, Montreal, QC, H3G 1M8
}

\maketitle

\begin{abstract}
  Graphs and hypergraphs are commonly distinguished by the order of the relations 
  they represent, with graphs encoding pairwise relations and hypergraphs encoding 
  interactions among arbitrary groups. We show that, for node-update dynamics, a more 
  natural viewpoint is to regard a graph as a collection of focal neighbourhoods, each centered 
  on a vertex and consisting of the vertex and its neighbours. From this 
  perspective, hypergraphs naturally generalise graph neighbourhoods by allowing 
  arbitrary interaction domains without requiring a distinguished focal node. This 
  viewpoint also clarifies the relationship between graph and hypergraph dynamical 
  models. Although sufficiently general graph-based node functions can reproduce the 
  dynamics of higher-order interactions, such dynamical equivalence does not imply 
  equivalence of the underlying interaction structures. Graph representations may 
  encode higher-order organisation implicitly in their functions while losing the 
  explicit grouping, roles, and symmetries carried by hyperedges. We therefore 
  distinguish dynamical expressivity from structural representation and show 
  that hypergraphs provide a generalisation of graph-based interaction 
  domains even when their dynamics can be emulated by graph models.
\end{abstract}

\section{Introduction}
What does it mean for a group of entities to interact?  In physics,
biology, and the social sciences alike, the answer shapes the
mathematical language one reaches for.  When interactions are binary,
graphs are the natural model.  When interactions are intrinsically
multi-body, the question of appropriate representation becomes more
subtle.

A substantial literature has advanced hypergraphs, simplicial complexes, and
related higher-order formalisms as natural languages for interactions that
cannot be reduced to pairwise relationships
\cite{Benson2016,Lambiotte2019,moutuou2023,Battiston2020,Torres2021}.
The distinction between graph and hypergraph models is nevertheless often left
implicit.  Our aim here is to make this distinction precise.  The key, we
argue, is that it depends on what is meant by an \emph{interaction}.  In
particular, two levels of description should be kept carefully apart:
\begin{itemize}
\item the \emph{structure} of a network, which records how entities are grouped
or connected; and
\item the \emph{dynamics} that runs on it, which governs how the state of each
entity evolves.
\end{itemize}
A bridge between two landmasses exists as a structural fact independently of
any traffic that crosses it.  At the dynamical level, however, the flow through
a road junction may depend simultaneously on all roads meeting there.  The
elementary structural relation and the domain of the dynamical interaction
therefore need not coincide.  Making this distinction explicit is essential to
a precise comparison between graph and hypergraph models.

For node-update dynamics on a graph, this distinction leads to a simple but
consequential observation.  Although a graph is structurally defined by
pairwise relations, the natural domain of a node's dynamics is generally its
closed neighbourhood rather than an individual edge.  If the state of node
$i$ evolves according to
\begin{equation}
\dot{x}_i=F_i\bigl(x_i,\{x_j:j\in\partial i\}\bigr),
\label{eq:intro-neighbourhood}
\end{equation}
then $F_i$ may depend jointly and nonlinearly on all states in the closed
neighbourhood $N[i]$, with $i$ serving as its distinguished \emph{focal}
node.  This observation underlies recent arguments that graph neighbourhood
functions can already encode higher-order dynamics \cite{Peixoto2026}, but it
also suggests a different interpretation of what such a graph model
represents. From the viewpoint of its dynamical interaction domains, the graph
admits a canonical \emph{focal neighbourhood representation}: its closed
neighbourhoods are interaction domains equipped with their corresponding focal
nodes.  The representation is canonically equivalent to the graph and therefore
adds no structural information to it.  Its role is to place graph neighbourhood
models and general hypergraph models in a common interaction domain language.
This separates two comparisons that otherwise become easily conflated; \emph{i.e.}, edges
and hyperedges as structural relations, and pointed neighbourhoods and
hyperedges as domains of dynamical functions. General hypergraphs can then be
viewed as extending graph-derived neighbourhood systems by allowing arbitrary
indexed group domains beyond those induced by graph adjacency.

\medskip

The paper develops this viewpoint in several steps.  We first introduce a 
taxonomy of structural, focal, and non-focal interaction domains, formulated in 
terms of indexed interaction supports so that repeated events and multiple focal 
nodes can be represented explicitly.  We then distinguish the conventional structural 
embedding of a graph as a $2$-uniform hypergraph from its dynamical interpretation as 
a family of pointed neighbourhoods, and show that this focal neighbourhood representation 
is canonically equivalent to the underlying graph. The purpose of this correspondence is 
to provide a common interaction-domain language in which graph and hypergraph dynamics 
can be compared independently of their particular representation. Within this language, we 
characterise the structural extension from graph-derived neighbourhood systems to general 
hypergraphs, distinguish edge-factorised from neighbourhood-domain graph dynamics, and 
separate these structural questions from dynamical realisability. This distinction provides the basis for revisiting recent maximal expressivity claims for 
graph models~\cite{Peixoto2026} and for identifying what is actually gained, and what is merely re-encoded, 
when passing between graph and hypergraph descriptions.

\medskip

The practical upshot is a principle of representational alignment.  Specifically, dynamical
expressivity alone does not determine whether a graph or hypergraph is the more
appropriate representation.  If pairwise relations and edge factorised
mechanisms are the objects of interest, the graph edge representation is the
natural one.  If a graph specifies the variables entering each node update, its
interaction domains are naturally focal neighbourhoods.  If instead group event
identity, multiplicity, source--target organisation, symmetry, or shared
interaction level constraints are part of the phenomenon to be represented, a
hypergraph keeps that information explicit rather than encoding it implicitly
in node functions or auxiliary structure.  Conversely, when only the realised
vector field matters, graph, focal, and non-focal parameterisations may be
dynamically interchangeable.  The relevant question is therefore not simply
which formalism can reproduce a given dynamics, but which interaction structure
the model is intended to represent.

The paper is organised as follows.  Section~\ref{sec:taxonomy} introduces the
interaction-domain taxonomy.  Section~\ref{sec:interaction-systems} defines
indexed focal and non-focal systems.  Section~\ref{sec:embeddings} distinguishes
the structural edge embedding from the dynamical neighbourhood embedding. Section~\ref{sec:graphmodels}
distinguishes two classes of graph dynamics.  Section~\ref{sec:hypergraphmodels}
defines hypergraph dynamics and non-focal symmetry.  Section~\ref{sec:projection}
analyses projection, behavioural compilation, and focal decomposition.
Section~\ref{sec:example} gives a worked comparison, and
Sections~\ref{sec:universal}--\ref{sec:discussion} discuss universal encodings
and the implications for the graph--hypergraph debate.

\section{A Taxonomy of interaction types}
\label{sec:taxonomy}

Before developing the formal concepts, it is useful to distinguish three
types of interaction that arise in complex systems and that make
different demands on the representational language.  The taxonomy is not
a partition of phenomena (real systems may involve all three
types simultaneously) but a partition of the questions one can ask.

\begin{enumerate}
  \item \textbf{Structural--Topological (ST) interaction.}\quad A binary
    combinatorial relation between two entities, independent of any
    dynamical state.  A bridge, a synaptic connection, a territorial
    boundary.  These are pairwise by nature and are faithfully
    represented by graph edges.  The existence of the bridge is a
    structural fact that does not depend on whether any flow is present.

  \item \textbf{Focal--Dynamical (FD) interaction.}\quad A process whose interaction domain is organised relative to a designated
  reference node or reference set.  The canonical example is a node-update
  equation on a graph: the state of vertex $i$ evolves as a function of the
  states in $\N{i}$, with $i$ identifying the equation to which the domain
  belongs.  The dependence may be irreducibly multivariate; focality concerns
  the organisation of the domain, not pairwise additivity.
  
  \item \textbf{Non-Focal (NF) group interaction.}\quad An indexed interaction event 
  represented without a designated reference
  participant.  The group, rather than one member's neighbourhood, is the
  primitive event.  If the participants are exchangeable, the associated
  interaction functional is naturally permutation equivariant.  If they have
  typed roles, the relevant symmetry may be a proper subgroup or there may be
  no nontrivial permutation symmetry at all.
\end{enumerate}

The FD/NF distinction is therefore independent of cardinality.  A large graph neighbourhood 
is focal because it is the neighbourhood of a
specified vertex.  A two-member interaction can be non-focal if it is recorded
as one unpointed joint event.  Conversely, a three-member interaction can be
focal, multifocal, or non-focal depending on its role structure.

\begin{remark}[Focality is not causality]
Notice that a focal node need not be the unique causal source, and it need not be the only
updated participant.  Focality records a distinguished reference role in the
interaction domain representation.  Directed applications may separately
specify source and target sets.  This distinction is important for real systems
such as a transcription networks, where the transcription factor
is a natural focal source while several target genes receive the update.
\end{remark}

\section{Indexed focal and non-focal interaction systems}
\label{sec:interaction-systems}

Recall~\cite{Berge1976} that a \emph{hypergraph} is a pair $H = (V, E)$ where $V$ is a
finite set of vertices and $E \subseteq \mathcal{P}(V) \setminus
\{\emptyset\}$ is a collection of non-empty subsets of $V$ called
\emph{hyperedges}. Observe that such a set-valued hyperedge collection is 
insufficient when distinct interaction occurrences have the same support, which occurs 
naturally in temporal data, multihypergraphs, and even in graph neighbourhood representations 
(\emph{e.g.}, distinct vertices of a complete graph have the same closed neighbourhood). We 
therefore need the following:

\begin{definition}[Indexed hypergraph]
An \emph{indexed hypergraph} is a triple 
\begin{equation}
  \calH=(V,\calI,s),
\end{equation}
where $V$ is a finite vertex set, $\calI$ is a finite set of interaction
occurrences (\emph{indexed hyperedges}), and
\begin{equation}
  s:\calI\longrightarrow \calP(V)\setminus\{\varnothing\}
\end{equation}
assigns a non-empty support to each occurrence, where $\calP(V)$ is the the collection of all subsets of $V$.  The map $s$ need not be
injective. 
\end{definition}

An ordinary simple hypergraph is recovered when $s$ is injective and the
interaction indices are identified with their supports.  The indexed
formulation retains multiplicity and permits several occurrences with the same
participants.

\begin{definition}[Focal system]
  A \emph{focal interaction system} is a quadruple
\begin{equation}
  \calF=(V,\calI,s,\varphi),
\end{equation}
where $(V,\calI,s)$ is an indexed hypergraph and
\begin{equation}
  \varphi:\calI\longrightarrow \calP(V),
  \qquad \varphi(a)\subseteq s(a)
\end{equation}
assigns a possibly empty \emph{focal set} to each occurrence.
\end{definition}

An occurrence $a$ is:
\begin{itemize}
  \item \emph{unifocal} if $|\varphi(a)|=1$;
  \item \emph{multifocal} if $|\varphi(a)| > 1$;
  \item \emph{non-focal} if $\varphi(a)=\varnothing$.
\end{itemize}

The term ``focal'' is chosen by analogy with
the focal point of a lens or the focus of an ellipse; \emph{i.e.}, a distinguished
centre relative to which the surrounding structure is defined.

\subsection{Examples of focal systems}

\paragraph{Graph neighbourhood update.}
For a graph model, the occurrence associated with vertex $i$ has support
$\N{i}$ and singleton focal set $\{i\}$.

\paragraph{Leader and constituency (social).}
A leader and constituency may define a focal social domain when the relation is
organised around the role holder.  If a fanbase supports an indivisible group
of athletes (\emph{e.g.}, a football team) rather than one athlete, the natural representation is multifocal,
with the athlete group as the focal set.  The indexed definition permits
several occurrences with identical constituent sets and different focal sets.

\paragraph{Transcription factor and its target gene set (biology).}
A transcription factor and its target genes form a directed one-to-many
interaction domain \cite{ShenOrr2002,Alon2006,Alon2007}.  The transcription
factor is a natural focal source, whereas the target set may contain several
updated nodes.  This example motivates distinguishing the focal set from the
output coordinates of the dynamical functional.

\paragraph{Server and connected clients (distributed systems).}
A server $S$ maintains state as a function of a set of currently
connected clients $\{C_1, \ldots, C_k\}$.  The hyperedge $(\{S, C_1,
\ldots, C_k\}, S)$ is focal at $S$: the server's load, cache, session
state, and response policy are all functions of its client
neighbourhood.  The clients do not interact with one another through
this hyperedge; they are the context within which $S$ operates.

\paragraph{Keystone species and its dependents (ecology).}
A keystone species $K$ sustains a community of species $\{D_1, \ldots,
D_k\}$ that depend on it for resources, regulation, or habitat.  The
concept was introduced by Paine~\cite{Paine1966,Paine1969} through
experiments showing that removing the predatory sea star
\textit{Pisaster ochraceus} from an intertidal community caused a
collapse from co-existing species to a mussel monoculture: the
entire community was organised around the focal species.  The hyperedge
$(\{K, D_1, \ldots, D_k\}, K)$ is focal at $K$: remove $K$ and the
community collapses; remove one $D_i$ and only that dependent's
relationship to $K$ changes.

\subsection{Examples of non-focal systems}
The following systems have the structure of a non-focal system
\emph{intrinsically}; namely, the interaction belongs to the group as a whole,
and asking which members are focal node receives no coherent answer.
In each case the group interaction cannot, even in principle, be
reconstructed from focal sub-interactions without information loss.

\paragraph{Irreducible three-body interaction (nuclear physics).}
A three-nucleon interaction is naturally recorded as one group event rather
than three independently observed focal events \cite{Fujita1957}.  Its equations
of motion can of course be written coordinate by coordinate; non-focality here
concerns the organisation of the physical interaction term as one shared event.

\paragraph{Higher-order epistasis (genetics).}
A fitness contribution assigned to a combination of mutations beyond all lower
order contributions is naturally indexed by the mutation set as a whole
\cite{Sailer2017,Weinreich2018}.  No mutation need be selected as the reference
member of the interaction event.

\paragraph{Multi-species coexistence via higher-order interactions (ecology).}
In competitive ecological communities, three or more species can coexist
stably even when every pair would be driven to competitive exclusion
under purely pairwise dynamics.  The stabilising force is a genuinely
three-way interaction: the presence of species $C$ modifies the
interaction between $A$ and $B$ in a way that cannot be decomposed into
any combination of focal bilateral effects.  Grilli et
al.~\cite{Grilli2017} showed analytically and numerically that
higher-order interactions of this kind stabilise diverse competitive
communities, with stability increasing with community size (the
opposite of what pairwise models predict).  The interaction $\{A,B,C\}$
is non-focal: representing it as a graph forces one species into the
role of focal node, misrepresenting a symmetric group context as a
bilateral influence, and destroys the stabilising structure.

\paragraph{Social norms and correlated equilibria (social and game theory).}
A correlated equilibrium is a joint distribution over strategy profiles, not
in general a product of independent marginals \cite{Aumann1974,Aumann1987}.
Player-centred incentive inequalities can be derived from it, but the common
coordination device remains a group-level object.

\section{Structural and dynamical embeddings}
\label{sec:embeddings}

There are two natural ways to
embed a graph into hypergraph theory, and the confusion between
them is the source of apparent conflicts in the literature.

\paragraph{The structural embedding.}
Combinatorially, a simple graph $G = (V,E)$ is a ($2$--uniform) hypergraph in which
every hyperedge has cardinality~2.  This is the conventional \emph{structural
embedding}; it characterises a graph by its elementary building blocks
(binary relations) and generalises by relaxing the cardinality
constraint. This is the view implicit in most of the higher-order
networks literature when it says ``hypergraphs generalise graphs.''  It
corresponds to the structural level of description where edges are the
objects, and hyperedges generalise them.

\begin{remark}
The structural edge embedding is appropriate when the pairwise relations are
the data or when the dynamics is factorised into edge-indexed mechanisms.  In
that setting, replacing edges by closed neighbourhoods without retaining the
edge decomposition would hide relevant structure.
\end{remark}

\paragraph{The dynamical neighbourhood embedding.}
For the neighbourhood domain model \eqref{eq:intro-neighbourhood}, the function
associated with vertex $i$ is evaluated on the states in $\N{i}$.  The domain is
therefore the pointed set $(\N{i},i)$, not an individual edge. For instance, the traffic flow through a junction depends on all
roads meeting there; the membrane potential of a neuron integrates all
its pre-synaptic inputs; opinion dynamics at node $i$ responds to its
entire social context.  

Formally, this gives the correspondence
\begin{equation}
  \Nf:G\longmapsto
  \bigl(V,V,s_G,\varphi_G\bigr),
  \qquad
  s_G(i)=\N{i},\quad \varphi_G(i)=\{i\}.
  \label{eq:focal-neighbourhood-map}
\end{equation}
We refer to the system $\Nf(G)$ obtained from \eqref{eq:focal-neighbourhood-map} as the \emph{focal neighbourhood
representation} of $G$.
Whereas the structural embedding identifies edges with $2$-element
hyperedges, $\Nf(G)$ organises the same graph by the
pointed domains on which its node-update functions act.

\begin{remark}
The apparent conflict identified by Peixoto et al.~\cite{Peixoto2026}
arises from comparing graph dynamical models to hypergraph models
under the structural embedding. Under the dynamical embedding, which is the natural frame for
comparing dynamical models, graph neighbourhoods and hyperedges are
objects of the same type (focal and non-focal interaction domains
respectively), and the symmetry condition is not an additional
restriction but the definition of a non-focal interaction domain. 
\end{remark}

In fact, the relation between the graph and focal domains can be made more precise by the following 
straightforaward correspondence.

\begin{lemma}[Graph--focal-neighbourhood correspondence]
\label{lem:graph-neighbourhood}
For a finite vertex set $V$, simple undirected graphs on $V$ are in
canonical bijection with indexed unifocal systems $\{(D_i,i)\}_{i\in V}$
satisfying
\begin{align}
  i &\in D_i
  &&\text{for every }i\in V,
  \label{eq:reflexivity}\\
  j\in D_i &\Longleftrightarrow i\in D_j
  &&\text{for every }i,j\in V.
  \label{eq:reciprocity}
\end{align}
The correspondence is given by \(D_i=N_G[i]\), with inverse
\begin{equation}
  \{i,j\}\in E(G)
  \quad\Longleftrightarrow\quad
  i\neq j\ \text{and}\ j\in D_i .
  \label{eq:recover-graph}
\end{equation}
\end{lemma}

\begin{proof}
For $D_i=N_G[i]$, reflexivity follows from the definition of the closed
neighbourhood and reciprocity from symmetry of adjacency.  Conversely,
\eqref{eq:recover-graph} defines an irreflexive symmetric adjacency relation
whenever \eqref{eq:reflexivity}--\eqref{eq:reciprocity} hold, and its closed
neighbourhoods satisfy
\[
  N_G[i]
  =
  \{i\}\cup\{j\neq i:j\in D_i\}
  =
  D_i.
\]
Hence the two constructions are mutually inverse.
\end{proof}

Observe that, as mentioned earlier, interaction indices are needed because different focal domains may have the
same support.  In particular, for the complete undirected graph $K_n$, $N[i]=V$ for every $i$, while the
indexed representation retains the $n$ distinct occurrences $(V,i)$.

Furthermore, forgetting the focal markers gives the indexed neighbourhood hypergraph
\begin{equation}
  \mathcal N(G)
  =
  \bigl(V,V,i\mapsto N[i]\bigr).
  \label{eq:neighbourhood-hypergraph}
\end{equation}
Such hypergraphs satisfy constraints inherited from graph adjacency; namely, their
interaction occurrences are indexed by vertices, each support contains its
indexing vertex, and membership is reciprocal.  On the other hand, general indexed hypergraphs
relax these constraints, allowing arbitrary supports, repeated occurrences,
and overlap patterns that need not arise from graph neighbourhoods; focal
information may likewise be absent or take more general forms.  Hence general
hypergraphs strictly extend graph-derived neighbourhood systems at the level of
interaction domain structure, independently of whether their dynamics can be
realised by sufficiently general graph neighbourhood functions.

\begin{remark}
For directed graphs, the same construction applies to in-neighbourhoods,
out-neighbourhoods, or both, with reciprocity replaced by the corresponding
directed incidence condition.
\end{remark}

\section{Two classes of graph dynamical models}
\label{sec:graphmodels}

The term ``\emph{graph dynamical model}'' is used for two distinct functional
architectures.  Their separation is necessary for the expressivity question.

\subsection{Pairwise dynamics}
In edge factorised model, each graph edge indexes a separate interaction term. A standard continuous time form is 
\begin{equation}
  \dot x_i
  =f_i(x_i)+\sum_{j:\{i,j\}\in E}g_{ij}(x_i,x_j).
  \label{eq:pairwise-graph}
\end{equation}
Here the graph records both the possible dependencies and their pairwise factorisation. 

One can see that this case does not contain every multivariate neighbourhood dependence on a fixed state space. Indeed, suppose $j$ and $k$ are distinct neighbours of $i$. Then, the term 
\begin{equation}
  \dot x_i=x_jx_k
  \label{eq:mixed-term}
\end{equation}
cannot be written on an open domain as
$g_{ij}(x_i,x_j)+g_{ik}(x_i,x_k)$ with twice continuously differentiable
functions, because
\begin{equation}
  \frac{\partial^2}{\partial x_j\partial x_k}(x_jx_k)=1,
\end{equation}
whereas the mixed derivative of an edge-additive expression is zero.  Therefore,
pairwise edge-factorised dynamics are strictly restricted as function classes.

\subsection{Neighbourhood domain dynamics}
In the broader class used by Peixoto et al.~\cite{Peixoto2026}, the graph specifies only the domain of each node function:
\begin{equation}
  \dot x_i=F_i\bigl(\pi_{\N{i}}x\bigr),
  \label{eq:neighbourhood-graph}
\end{equation}
where $\pi_{\N{i}}$ projects onto the coordinates in $\N{i}$.  The function $F_i$ may be arbitrary and need not decompose into edge terms.  Equation \eqref{eq:mixed-term} is therefore allowed whenever $j,k\in\N{i}$.

Clearly, every edge factorised model \eqref{eq:pairwise-graph} is a neighbourhood domain model, obtained by
setting
\begin{equation}
  F_i(x_{\N{i}})=f_i(x_i)+\sum_{j:\{i,j\}\in E}g_{ij}(x_i,x_j).
\end{equation}
The converse is not true, in general. Hence the statement that graph models are or
are not restricted to pairwise interactions depends on which model class is meant.

\begin{remark}[Adjacency and factorisation]
  The focal neighbourhood family preserves the graph adjacency through Lemma~\ref{lem:graph-neighbourhood}. Nevertheless, replacing the edge indexed functions $g_{ij}$ by one unrestricted function $F_i$ does not preserve the edgewise factorisation of the dynamics unless that decomposition is stored separately. In other words, recoverability of the graph does not imply recoverability of its pairwise dynamical mechanisms.
\end{remark}

\section{Hypergraph dynamics and non-focal symmetry}
\label{sec:hypergraphmodels}
Let $\calH=(V,\calI,s)$ be an indexed hypergraph and let $n=|V|$.  For each occurrence $a\in\calI$, let
\begin{equation}
  \pi_a:\R^n\longrightarrow\R^{|s(a)|}
\end{equation}
be the coordinate projection onto $s(a)$, and let
\begin{equation}
  \iota_a:\R^{|s(a)|}\longrightarrow\R^n
\end{equation}
be the zero-padding embedding.  An \emph{interaction functional} is a map
\begin{equation}
  \Psi_a:\R^{|s(a)|}\longrightarrow\R^{|s(a)|}.
\end{equation}
The coordinate $\Psi_{a,i}$ gives the contribution of occurrence $a$ to the
rate of change of $i\in s(a)$.  The resulting hypergraph dynamical system~\cite{Carletti2020,Mulas2020,Bick2023} is
\begin{equation}
  \dot x
  =\sum_{a\in\calI}\iota_a\Psi_a(\pi_a x),
  \qquad
  \dot x_i
  =\sum_{a:\,i\in s(a)}\Psi_{a,i}(\pi_a x).
  \label{eq:general-hypergraph}
\end{equation}
Repeated supports remain distinct because the sum is indexed by $\calI$.

\subsection{Graph models as focal hypergraph dynamics}

For a graph $G$, take the focal neighbourhood system
$\calN_{\mathrm f}(G)$ and define, for the occurrence indexed by $i$,
\begin{equation}
  \Psi_{i,j}(\pi_i x)=
  \begin{cases}
    F_i(\pi_{\N{i}}x),&j=i,\\
    0,&j\neq i.
  \end{cases}
  \label{eq:graph-as-focal}
\end{equation}
Then \eqref{eq:general-hypergraph} reduces exactly to
\eqref{eq:neighbourhood-graph}.  Thus neighbourhood domain graph dynamics are dynamics on the focal neighbourhood hypergraph.

This formulation makes the graph constraints more explicit; namely, there is one occurrence
per vertex, its support is the graph-derived neighbourhood, and its output is
assigned to the focal coordinate. General focal systems may relax all three
conditions.

\subsection{Non-focality and equivariance}
Non-focality is, fundamentally, a structural statement defined by $\varphi(a)=\varnothing$. Being non-focal does not automatically imply that all participants are physically identical. However, when the participants of a hyperedge $a$ are interchangeable, the appropriate dynamical symmetry is permutation equivariance. Specifically, in such a case, letting $P_\sigma$ denote the permutation
matrix associated with $\sigma\in S_{|s(a)|}$, we get
\begin{equation}
  \Psi_a(P_\sigma y)=P_\sigma\Psi_a(y)
  \qquad
  \text{for every }\sigma\in S_{|s(a)|}.
  \label{eq:equivariance}
\end{equation}
More generally, the full symmetric group may be replaced by a role-preserving subgroup $\Gamma_a$.

Importantly, equivariance must not be confused with identical output coordinates. For
example, the function
\begin{equation}
  \Psi_{a,i}(y)
  =\frac{1}{|s(a)|}\sum_{j\in s(a)}y_j-y_i,
  \label{eq:consensus-functional}
\end{equation}
for $|s(a)|> 1$, is permutation equivariant but gives different coordinate values whenever the
states differ.  No labelled vertex is privileged; \emph{i.e.}, relabelling the participants
only relabels their responses.

\begin{remark}
A non-focal interaction may also have directed or typed internal roles~\cite{Lotito2026}.  In
such cases the relevant primitive is the common indexed event together with its
source--target partition or role labels, rather than full permutation
symmetry.
\end{remark}

\section{Projection, behavioural compilation, and focal decomposition}
\label{sec:projection}
Observe that the three operations of projection, dynamical compilation, and focal decomposition affect different aspects of an interaction model. We consider them separately to distinguish changes in interaction structure from changes in the representation of a vector field.

\subsection{Projection and compilation}
For an indexed hypergraph $\calH=(V,\calI,s)$, let $G=\TwoSec(\calH)$ denote its
\emph{$2$-section graph}, defined by
\begin{equation}
  \{i,j\}\in E(G)
  \quad\Longleftrightarrow\quad
  i\neq j
  \text{ and }
  \{i,j\}\subseteq s(a)
  \text{ for some }a\in\calI .
  \label{eq:2section}
\end{equation}
The closed neighbourhood of $i$ is then
\begin{equation}
  N_G[i]
  =
  \bigcup_{a:\,i\in s(a)}s(a).
  \label{eq:union-neighbourhood}
\end{equation}
Hence $N_G[i]$ records all vertices that co-occur with $i$, while the
division of these vertices among individual interaction occurrences is lost
under projection.  For example,
\begin{equation}
  \calH_1=\bigl\{\{1,2,3\}\bigr\},
  \qquad
  \calH_2=
  \bigl\{\{1,2\},\{1,3\},\{2,3\}\bigr\}
  \label{eq:triangle-hypergraphs}
\end{equation}
both project to $K_3$, although one records a single three-vertex occurrence
and the other three pairwise occurrences.  More generally, multiplicities and
other indexed collections may produce the same projection.  Recovering such
higher-order structure from graph data is consequently an inverse problem
\cite{Young2021,Wang2024}. Moreover, the loss of interaction structure under projection remains compatible with
exact reproduction of the associated dynamics.

\begin{proposition}[Behavioural compilation]
\label{prop:behavioural-compilation}
Let $\calH=(V,\calI,s)$ carry the vector field
\eqref{eq:general-hypergraph}, and let $G=\TwoSec(\calH)$.  For every
node $i\in V$, there exists a function
\[
  F_i:\mathbb R^{|N_G[i]|}\longrightarrow\mathbb R
\]
such that
\begin{equation}
  \dot x_i=F_i(\pi_{N_G[i]}x).
  \label{eq:compiled-graph}
\end{equation}
\end{proposition}

\begin{proof}
For every occurrence $a$ containing $i$,
\eqref{eq:union-neighbourhood} gives $s(a)\subseteq N_G[i]$.  Define
\begin{equation}
  F_i(z)
  =
  \sum_{a:\,i\in s(a)}
  \Psi_{a,i}\bigl(\pi_{s(a)}z\bigr).
  \label{eq:compiled-function}
\end{equation}
Setting $z=\pi_{N_G[i]}x$ recovers the $i^{th}$ coordinate of
\eqref{eq:general-hypergraph}.
\end{proof}

Proposition~\ref{prop:behavioural-compilation} establishes exact dynamical
realisability on the projected graph for the model class considered here.  The
construction of \(F_i\) nevertheless uses the original occurrences \(a\), their
supports \(s(a)\), and the associated functionals \(\Psi_{a,i}\), none of which
is determined in general by \(G\).  Distinct hypergraph structures may
therefore yield the same projected graph and admit graph neighbourhood
functions reproducing their respective vector fields.  It follows that the expressive object
in this construction is the graph together with its node functions,
\begin{equation}
  \bigl(G,\{F_i\}_{i\in V}\bigr),
\end{equation}
while the graph specifies only their neighbourhood domains.

\subsection{Focal decomposition}
A related equivalence arises at the level of individual interaction
functionals.  For each occurrence $a\in\calI$ and each
$i\in s(a)$, define the focal component
\begin{equation}
  \Psi^{(i)}_{a,j}(y)
  =
  \begin{cases}
    \Psi_{a,i}(y), & j=i,\\
    0,             & j\neq i .
  \end{cases}
  \label{eq:focal-component}
\end{equation}

\begin{proposition}[Focal decomposition]
\label{prop:focal-decomposition}
Every interaction functional $\Psi_a$ admits the decomposition
\begin{equation}
  \Psi_a(y)
  =
  \sum_{i\in s(a)}\Psi^{(i)}_a(y),
  \label{eq:focal-sum}
\end{equation}
and replacing $a$ by the focal occurrences
$\{(a,i):i\in s(a)\}$ preserves the global vector field.
\end{proposition}

\begin{proof}
For any $j\in s(a)$, the $j^{th}$ coordinate of
\eqref{eq:focal-sum} receives a contribution only from the term $i=j$,
which equals $\Psi_{a,j}(y)$.  The equality therefore holds
coordinatewise, and substitution into \eqref{eq:general-hypergraph} leaves
the vector field unchanged.
\end{proof}

The decomposition shows that unrestricted vector field realisability does not
separate focal from non-focal parameterisations.  Their distinction instead
concerns the organisation assigned to the interaction occurrence.  Replacing a
single vector-valued event by separately indexed focal components can remove
from the primitive representation its common event identity, multiplicity,
permutation or role equivariance, source--target organisation, shared
parameters, and interaction-level constraints.  These features remain
available when the focal components retain their parent occurrence and the
corresponding constraints, but otherwise become properties imposed on the
decomposition rather than properties encoded by a single interaction event.
Thus focal decomposition preserves the vector field while potentially changing
its interaction factorisation.

\section{A toy example: one triple or three pairs?}
\label{sec:example}
The two hypergraphs $\calH_1$ and $\calH_2$ defined in \eqref{eq:triangle-hypergraphs} illustrate the entire issue in a
minimal form.  Both project to the graph
\begin{center}
\begin{tikzpicture}[
  every node/.style={circle,draw,minimum size=8mm},
  every path/.style={line width=0.7pt}]
  \node (1) at (0,1.5) {$1$};
  \node (2) at (-1.3,-0.6) {$2$};
  \node (3) at (1.3,-0.6) {$3$};
  \draw (1)--(2)--(3)--(1);
\end{tikzpicture}
\end{center}

Now, for $\calH_1$, suppose the triple carries one equivariant interaction functional
\begin{equation}
  \Psi_i(x_1,x_2,x_3)
  =\frac{x_1+x_2+x_3}{3}-x_i,
  \qquad i=1,2,3.
  \label{eq:triple-consensus}
\end{equation}
This is one interaction occurrence, one parameterised functional, and one
symmetry constraint.

For $\calH_2$, suppose instead that the three edges carry independent pairwise
functions.  The structural assumptions differ even if the resulting vector
field is tuned to coincide with \eqref{eq:triple-consensus}.  An intervention
that removes the triple in $\calH_1$ removes the whole event.  In $\calH_2$,
one pair can be removed while the other two remain.  Repeated triple events can
be counted in $\calH_1$, whereas their multiplicity is invisible in $K_3$.

On the other hand, the projected graph permits arbitrary functions
\begin{equation}
  \dot x_i=F_i(x_1,x_2,x_3),
\end{equation}
so it can reproduce either system.  But the graph alone does not determine
which event structure the functions encode. It follows that calling the graph ``maximally
expressive'' is correct only in the extensional sense that the
functions can realise both vector fields.  It is, however, not correct if maximal
expressivity is taken to mean that $K_3$ contains or identifies both interaction
structures.

Of note, the same graph also allows the irreducible focal term
\begin{equation}
  \dot x_1=x_2x_3,
\end{equation}
which, as shown earlier, cannot be represented by an additive sum of twice differentiable edge
functions on the fixed state space.  This illustrates the second distinction that
neighbourhood domain graph dynamics are broader than pairwise edge factorised
graph dynamics, even though both use the same graph support.

\section{Universal encodings and what they establish}
\label{sec:universal}
Every indexed hypergraph admits a bipartite incidence graph representation.
Specifically, given $\calH=(V,\calI,s)$, introduce a factor node $f_a$ for each
interaction occurrence $a\in\calI$ and connect $f_a$ to every vertex
in $s(a)$~\cite{Berge1976,Peixoto2026}.  The resulting bipartite graph has
vertex set
\[
  V\sqcup\{f_a:a\in\calI\},
\]
so each interaction occurrence remains individually represented.  Repeated
supports are distinguished by their factor nodes, while focal sets,
source--target roles, orientations, or symmetry data can be carried by labels
or typed incidences.

This incidence representation differs fundamentally from the projection
considered in Section~\ref{sec:projection}.  A clique projection retains the
original vertex set and records co-occurrence through adjacency, generally
discarding the identity of the interaction occurrences that generated it.
An incidence graph preserves those occurrences by representing each of them as
an additional node.  Hyperedge identity is therefore transferred from an
indexed family of supports to an auxiliary class of graph vertices.

The two constructions illustrate distinct senses in which higher-order
structure can be represented by a graph.  Projection seeks a graph on the
original entities and may lose interaction domain information, and incidence
expansion gives a faithful graph encoding by enlarging the representation.
More generally, graph encodings may preserve higher-order structure through
auxiliary vertices, labels, types, or constraints.  Their expressive power
therefore depends on the structure carried by the full encoding, while the
choice of primitive objects determines which aspects of the interaction
organisation remain explicit.

\section{Discussion}
\label{sec:discussion}
The results developed above support a distinction between the structural and
dynamical roles of a graph.  Structurally, a graph records pairwise adjacency.
In a neighbourhood domain dynamical model,
\begin{equation}
  \dot x_i=F_i(\pi_{N[i]}x),
\end{equation}
the function associated with vertex $i$ acts on the pointed domain
$(N[i],i)$.  The canonical correspondence established in
Section~\ref{sec:embeddings} therefore gives a focal hypergraph interpretation
of graph node-update dynamics; \emph{i.e.}, the interaction domains are indexed
neighbourhoods with their updating vertices as focal nodes.  This
interpretation preserves the underlying graph while making explicit the
domains on which its dynamical functions act. Its contribution is therefore 
conceptual rather than representational; \emph{i.e.}, it provides a common 
interaction domain language in which graph neighbourhood models and general 
hypergraph models can be compared while keeping structural organisation 
distinct from dynamical realisability.

General indexed hypergraphs extend these graph-derived domains by relaxing the
constraints inherited from adjacency.  Their interaction occurrences may have
arbitrary supports, repeated supports, more general overlap patterns, and focal
or non-focal organisation.  The extension concerns interaction structure
rather than unrestricted vector field realisability.  As shown in
Section~\ref{sec:projection}, hypergraph vector fields can be compiled into
functions on projected graph neighbourhoods, and vector-valued interactions can
be decomposed into focal coordinate contributions.  Consequently, graph,
focal, and non-focal formulations may generate the same vector field while
assigning different primitive structure to its interactions.

The distinction becomes consequential whenever interaction organisation forms
part of the model.  Projection records co-occurrence through graph adjacency
but may identify hypergraphs with different event groupings, multiplicities,
roles, or constraints.  Exact dynamical compilation then transfers the
information required to distinguish those systems into the node functions,
their parameter relations, or auxiliary structure.  Dynamical equivalence
therefore leaves open a separate representational question: which aspects of
the interaction organisation are explicit in the network object and which are
encoded in the accompanying functions or constraints?

\paragraph{Relation to Peixoto et al.}
The distinction developed here places the maximal expressivity argument of
Peixoto et al.~\cite{Peixoto2026} in a more precise setting.  Their mathematical
result establishes that graph neighbourhood functions need not be pairwise
additive and that, when sufficiently general functions are allowed, higher-order
vector fields can be reproduced on a graph representation.  This rules out a
strict hierarchy based solely on multivariate dependence or unrestricted
vector field realisability.

However, a stronger conclusion requires a different notion of expressivity.  In
particular, the claim that hypergraph models are strict special cases of graph
models, or that graphs are maximally expressive, is valid only if
``expressivity'' refers to the ability of a graph together with unrestricted
node functions to realise a given vector field.  The reduction does not
establish the corresponding statement for interaction structure.  Hypergraphs
that differ in their interaction occurrences, multiplicities, grouping, or
roles may have the same graph projection, while their respective dynamics are
recovered by different functions on that graph.  The information distinguishing
the original systems has then moved from the network structure into the
functions, their parameter constraints, or auxiliary variables.

This qualification changes the interpretation of the reduction; namely, the
universality result applies to the dynamical specification
\begin{equation}
  \bigl(G,\{F_i\}_{i\in V}\bigr),
\end{equation}
rather than to graph adjacency in isolation.  Consequently, behavioural
compilation shows that hypergraph structure need not enlarge the class of
vector fields obtainable from unrestricted graph neighbourhood functions; it
does not show that the graph carries all interaction structure represented by
the hypergraph. Calling hypergraphs ``special cases of graphs'' without this
qualification conflates inclusion between classes of dynamical realisations
with inclusion between classes of interaction domain representations.

The focal neighbourhood correspondence makes the source of this distinction
particularly transparent.  In the broad graph model used by Peixoto et al.,
the update at $i$ is already defined on the whole neighbourhood $N[i]$.
Its dynamical domain is therefore the pointed neighbourhood $(N[i],i)$,
rather than an individual graph edge.  Graph neighbourhood systems constitute
a constrained class of such focal domains, determined by a reciprocal
adjacency relation, whereas general hypergraphs permit interaction domains
with arbitrary indexed supports and, in the non-focal case, without a
distinguished reference node. The focal framework therefore contributes a 
common level of description on which the two claims can be separated: the expressive 
power of the dynamical functions concerns which vector fields can be realised, 
whereas the interaction domain representation concerns which organisation of the 
interactions is retained explicitly. The two conclusions can therefore be stated
without conflict: unrestricted graph neighbourhood functions may be maximally
expressive as vector field realisers, while general hypergraphs remain strictly
more general as explicit interaction domain representations.

\paragraph{Representational alignment.}
The choice of formalism should therefore follow the object one intends to
represent or infer.  Pairwise observations or edge factorised mechanisms are
naturally described by graph edges, whereas neighbourhood-based node dynamics
have a focal interaction domain interpretation.  When group event identity,
multiplicity, source--target organisation, event symmetry, or shared
interaction level parameters are empirically observed or mechanistically
meaningful, indexed hypergraphs retain these features directly.  When the
inferential target is only the realised vector field or its trajectories,
different graph, focal, and hypergraph parameterisations may instead be
equivalent after the appropriate functional constraints are imposed.
Representational adequacy consequently depends on both the dynamics of interest
and the structural information intended to remain explicit.


\bigskip
\noindent\textbf{Acknowledgements.}\quad We thank the complex systems
and higher-order networks communities for the discussions that motivated
this work.



\begin{thebibliography}{10}
\expandafter\ifx\csname url\endcsname\relax
  \def\url#1{\texttt{#1}}\fi
\expandafter\ifx\csname urlprefix\endcsname\relax\def\urlprefix{URL }\fi
\providecommand{\bibinfo}[2]{#2}
\providecommand{\eprint}[2][]{\url{#2}}

\bibitem{Benson2016}
\bibinfo{author}{Benson, A.~R.}, \bibinfo{author}{Gleich, D.~F.} \& \bibinfo{author}{Leskovec, J.}
\newblock \bibinfo{title}{Higher-order organization of complex networks}.
\newblock \emph{\bibinfo{journal}{Science}} \textbf{\bibinfo{volume}{353}}, \bibinfo{pages}{163--166}.

\bibitem{Lambiotte2019}
\bibinfo{author}{Lambiotte, R.}, \bibinfo{author}{Rosvall, M.} \& \bibinfo{author}{Scholtes, I.}
\newblock \bibinfo{title}{From networks to optimal higher-order models of complex systems}.
\newblock \emph{\bibinfo{journal}{Nature Physics}} \textbf{\bibinfo{volume}{15}}, \bibinfo{pages}{313--320}.

\bibitem{moutuou2023}
\bibinfo{author}{Moutuou, E.~M.}, \bibinfo{author}{Ali, O. B.~K.} \& \bibinfo{author}{Benali, H.}
\newblock \bibinfo{title}{Topology and spectral interconnectivities of higher-order multilayer networks}.
\newblock \emph{\bibinfo{journal}{Frontiers in Complex Systems}} \textbf{\bibinfo{volume}{1}} (\bibinfo{year}{2023}).
\newblock \urlprefix\url{https://www.frontiersin.org/articles/10.3389/fcpxs.2023.1281714}.

\bibitem{Battiston2020}
\bibinfo{author}{Battiston, F.} \emph{et~al.}
\newblock \bibinfo{title}{Networks beyond pairwise interactions: Structure and dynamics}.
\newblock \emph{\bibinfo{journal}{Physics Reports}} \textbf{\bibinfo{volume}{874}}, \bibinfo{pages}{1--92} (\bibinfo{year}{2020}).
\newblock \urlprefix\url{https://www.sciencedirect.com/science/article/pii/S0370157320302489}.
\newblock \bibinfo{note}{Networks beyond pairwise interactions: Structure and dynamics}.

\bibitem{Torres2021}
\bibinfo{author}{Torres, L.}, \bibinfo{author}{Blevins, A.~S.}, \bibinfo{author}{Bassett, D.} \& \bibinfo{author}{Eliassi-Rad, T.}
\newblock \bibinfo{title}{The why, how, and when of representations for complex systems}.
\newblock \emph{\bibinfo{journal}{SIAM Review}} \textbf{\bibinfo{volume}{63}}, \bibinfo{pages}{435--485}.

\bibitem{Peixoto2026}
\bibinfo{author}{Peixoto, T.~P.}, \bibinfo{author}{Peel, L.}, \bibinfo{author}{Gross, T.} \& \bibinfo{author}{De~Domenico, M.}
\newblock \bibinfo{title}{Graphs are maximally expressive for higher-order interactions}.
\newblock \emph{\bibinfo{journal}{arXiv preprint arXiv:2602.16937}} .

\bibitem{Berge1976}
\bibinfo{author}{Berge, C.}
\newblock \emph{\bibinfo{title}{Graphs and Hypergraphs}}.
\newblock North-Holland mathematical library (\bibinfo{publisher}{North-Holland Publishing Company}, \bibinfo{year}{1976}).
\newblock \urlprefix\url{https://books.google.ca/books?id=S59iNAEACAAJ}.

\bibitem{ShenOrr2002}
\bibinfo{author}{Shen-Orr, S.~S.}, \bibinfo{author}{Milo, R.}, \bibinfo{author}{Mangan, S.} \& \bibinfo{author}{Alon, U.}
\newblock \bibinfo{title}{Network motifs in the transcriptional regulation network of escherichia coli}.
\newblock \emph{\bibinfo{journal}{Nature Genetics}} \textbf{\bibinfo{volume}{31}}, \bibinfo{pages}{64--68}.

\bibitem{Alon2006}
\bibinfo{author}{Alon, U.}
\newblock \emph{\bibinfo{title}{An Introduction to Systems Biology: Design Principles of Biological Circuits}} (\bibinfo{publisher}{Chapman and Hall/CRC}).

\bibitem{Alon2007}
\bibinfo{author}{Alon, U.}
\newblock \bibinfo{title}{Network motifs: theory and experimental approaches}.
\newblock \emph{\bibinfo{journal}{Nature Reviews Genetics}} \textbf{\bibinfo{volume}{8}}, \bibinfo{pages}{450--461}.

\bibitem{Paine1966}
\bibinfo{author}{Paine, R.~T.}
\newblock \bibinfo{title}{Food web complexity and species diversity}.
\newblock \emph{\bibinfo{journal}{The American Naturalist}} \textbf{\bibinfo{volume}{100}}, \bibinfo{pages}{65--75} (\bibinfo{year}{1966}).
\newblock \urlprefix\url{http://www.jstor.org/stable/2459379}.

\bibitem{Paine1969}
\bibinfo{author}{Paine, R. T.~R.}
\newblock \bibinfo{title}{A note on trophic complexity and community stability}.
\newblock \emph{\bibinfo{journal}{The American Naturalist}} \textbf{\bibinfo{volume}{103}}, \bibinfo{pages}{91 -- 93} (\bibinfo{year}{1969}).
\newblock \urlprefix\url{https://api.semanticscholar.org/CorpusID:83780992}.

\bibitem{Fujita1957}
\bibinfo{author}{Fujita, J.-i.} \& \bibinfo{author}{Miyazawa, H.}
\newblock \bibinfo{title}{Pion theory of three-body forces}.
\newblock \emph{\bibinfo{journal}{Progress of Theoretical Physics}} \textbf{\bibinfo{volume}{17}}, \bibinfo{pages}{360--365}.

\bibitem{Sailer2017}
\bibinfo{author}{Sailer, Z.~R.} \& \bibinfo{author}{Harms, M.~J.}
\newblock \bibinfo{title}{High-order epistasis shapes evolutionary trajectories}.
\newblock \emph{\bibinfo{journal}{PLOS Computational Biology}} \textbf{\bibinfo{volume}{13}}, \bibinfo{pages}{e1005541}.

\bibitem{Weinreich2018}
\bibinfo{author}{Weinreich, D.~M.}, \bibinfo{author}{Lan, Y.}, \bibinfo{author}{Jaffe, J.} \& \bibinfo{author}{Heckendorn, R.~B.}
\newblock \bibinfo{title}{The influence of higher-order epistasis on biological fitness landscape topography}.
\newblock \emph{\bibinfo{journal}{Journal of Statistical Physics}} \textbf{\bibinfo{volume}{172}}, \bibinfo{pages}{208--225}.

\bibitem{Grilli2017}
\bibinfo{author}{Grilli, J.}, \bibinfo{author}{Barabás, G.}, \bibinfo{author}{Michalska-Smith, M.~J.} \& \bibinfo{author}{Allesina, S.}
\newblock \bibinfo{title}{Higher-order interactions stabilize dynamics in competitive network models}.
\newblock \emph{\bibinfo{journal}{Nature}} \textbf{\bibinfo{volume}{548}}, \bibinfo{pages}{210--213}.

\bibitem{Aumann1974}
\bibinfo{author}{Aumann, R.~J.}
\newblock \bibinfo{title}{Subjectivity and correlation in randomized strategies}.
\newblock \emph{\bibinfo{journal}{Journal of Mathematical Economics}} \textbf{\bibinfo{volume}{1}}, \bibinfo{pages}{67--96}.

\bibitem{Aumann1987}
\bibinfo{author}{Aumann, R.~J.}
\newblock \bibinfo{title}{Correlated equilibrium as an expression of bayesian rationality}.
\newblock \emph{\bibinfo{journal}{Econometrica}} \textbf{\bibinfo{volume}{55}}, \bibinfo{pages}{1}.

\bibitem{Carletti2020}
\bibinfo{author}{Carletti, T.}, \bibinfo{author}{Fanelli, D.} \& \bibinfo{author}{Nicoletti, S.}
\newblock \bibinfo{title}{Dynamical systems on hypergraphs}.
\newblock \emph{\bibinfo{journal}{Journal of Physics: Complexity}} \textbf{\bibinfo{volume}{1}}, \bibinfo{pages}{035006}.

\bibitem{Mulas2020}
\bibinfo{author}{Mulas, R.}, \bibinfo{author}{Kuehn, C.} \& \bibinfo{author}{Jost, J.}
\newblock \bibinfo{title}{Coupled dynamics on hypergraphs: Master stability of steady states and synchronization}.
\newblock \emph{\bibinfo{journal}{Physical Review E}} \textbf{\bibinfo{volume}{101}}, \bibinfo{pages}{062313}.

\bibitem{Bick2023}
\bibinfo{author}{Bick, C.}, \bibinfo{author}{Gross, E.}, \bibinfo{author}{Harrington, H.~A.} \& \bibinfo{author}{Schaub, M.~T.}
\newblock \bibinfo{title}{What are higher-order networks?}
\newblock \emph{\bibinfo{journal}{SIAM Review}} \textbf{\bibinfo{volume}{65}}, \bibinfo{pages}{686--731}.

\bibitem{Lotito2026}
\bibinfo{author}{Lotito, Q.~F.}, \bibinfo{author}{Betti, L.}, \bibinfo{author}{Nortier, B.}, \bibinfo{author}{Montresor, A.} \& \bibinfo{author}{Battiston, F.}
\newblock \bibinfo{title}{Hypergraphx-data: a repository for higher-order network data}.
\newblock \emph{\bibinfo{journal}{Journal of Complex Networks}} \textbf{\bibinfo{volume}{14}}.

\bibitem{Young2021}
\bibinfo{author}{Young, J.-G.}, \bibinfo{author}{Petri, G.} \& \bibinfo{author}{Peixoto, T.~P.}
\newblock \bibinfo{title}{Hypergraph reconstruction from network data}.
\newblock \emph{\bibinfo{journal}{Communications Physics}} \textbf{\bibinfo{volume}{4}}, \bibinfo{pages}{135} (\bibinfo{year}{2021}).

\bibitem{Wang2024}
\bibinfo{author}{Wang, Y.} \& \bibinfo{author}{Kleinberg, J.}
\newblock \bibinfo{title}{From graphs to hypergraphs: Hypergraph projection and its reconstruction}.
\newblock In \bibinfo{editor}{Kim, B.} \emph{et~al.} (eds.) \emph{\bibinfo{booktitle}{International Conference on Learning Representations}}, vol. \bibinfo{volume}{2024}, \bibinfo{pages}{52745--52775} (\bibinfo{year}{2024}).
\newblock \urlprefix\url{https://proceedings.iclr.cc/paper_files/paper/2024/file/e92b8441b702d142435f243806f9ec5a-Paper-Conference.pdf}.

\end{thebibliography}

\end{document}